\documentstyle[amssymb,prb,aps,epsfig]{revtex}

\begin{document}
\draft

\input{epsf}

\twocolumn[\hsize\textwidth\columnwidth\hsize\csname@twocolumnfalse\endcsname

\title{Kagom\'e spin ice}
\author{A.~S.~Wills\cite{ASW}, R.~Ballou,C.~Lacroix}

\address{Laboratoire  Louis N\'eel, CNRS, 25 Avenue des Martyrs, BP 166,
38042 Grenoble Cedex 9, France.}

\date{12 April 2002}
\maketitle

\begin{abstract}

A new model of localized highly frustrated ferromagnetism is
presented: {\it kagom\'{e}} spin ice. By use of analytical and
Monte Carlo calculations its massive groundstate entropy is
evaluated. Monte Carlo calculations are also used to explore the
phases in the presence of further-neighbor interactions and
applied fields. The importance of thermal spin fluctuations and
their ability to stabilize a variety of magnetic structures is
clearly manifested. Remarkably, some of these phases are only
partially ordered and present both disordered and ordered
sublattices.

\end{abstract}
\vspace{.2in}

\noindent ] 
\narrowtext

\section{Introduction}
\label{Introduction}

Geometric frustration refers to the inability of a magnetic system
to minimize its individual exchange interactions, as a simple
consequence of the exchange geometry. For many years it was thought to be limited to antiferromagnets,
with much research focusing on the {\it kagom\'e} and pyrochlore
antiferromagnets,\cite{Villain,Harris92,Ramirez94,Greedan01,HFM2000}
until a phase termed `spin ice' was discovered by Harris and
coworkers.\cite{spin_ice_Harris1,spin_ice_Harris2,spin_ice_Bramwell}
 In their
system, which is based on the pyrochlore lattice, the large number
of degenerate ground states that characterize frustrated magnetism
is a consequence of particular Ising anisotropies and
ferromagnetic nearest-neighbor exchange. Mapping the spin
directions onto the disordered hydrogen bonds of Pauling's cubic
ice model, their system was shown to remain disordered even at
T=0. However, if other terms are present, such as dipolar
interactions a transition to long-range order has been found due
to collective zero-energy flips of loops of spins.\cite{denHertog}
Despite the unique properties of pyrochlore spin ice, nothing is
known about 2-dimensional (2-d) spin ice models or more generally
about the properties of frustrated 2-d ferromagnetic systems. In contrast, 
the ice models were widely investigated in two dimensions, especially on a 
square lattice a version of which was solved exactly by Lieb. \cite{Lieb67,Baxter} 
Extensions and generalizations of the model and to other lattices soon followed.\cite{Baxter,LiebWu}

In this article we examine the single layer {\it kagom\'e}
ferromagnet with in-plane Ising anisotropies and identify a new spin ice
system--- {\it kagom\'e} spin ice. Using analytic and classical
Monte Carlo calculations we demonstrate the massive groundstate
degeneracy of this ice-like phase, which greatly exceeds that of
the pyrochlore spin ice. Furthermore, we show how its degeneracy
is raised by additional parameters in the Hamiltonian, namely
second neighbor interactions or an applied magnetic field, and that
these lead to the stabilization of a number of exotic magnetic
phases.

\begin{figure}[tbp]
\centerline{\hbox{\epsfig{figure=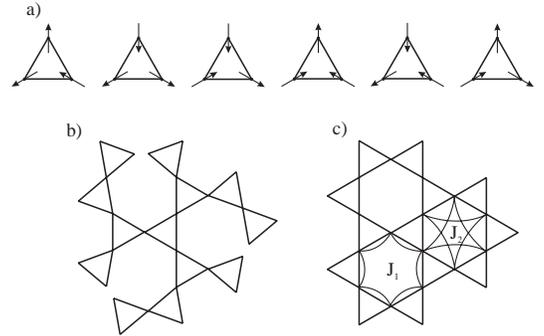,width=8cm}}}
\caption{a) The 6 degenerate spin configurations for a single
triangle of the {\it kagom\'e} spin ice model with the Ising axes
pointing to the center of each triangle. The Bethe lattice b) and the
{\it kagom\'e} lattice c) of vertex sharing triangles. The
nearest-neighbor and second neighbor interactions, $J_1$ and
$J_2$ respectively, of the {\it kagom\'e} lattice are defined
separately in c).} \label{lattices_figures}
\end{figure}

\section{The ice rules}
\label{The ice rules}

These local conditions
define whether a spin configuration lies within the ground state
manifold. For the pyrochlore system,\cite{spin_ice_Harris1,spin_ice_Harris2,spin_ice_Bramwell} where the Ising axes point
towards the centre of the basic tetrahedral motif, the ice rules are that
2 spins enter and 2 spins leave each tetrahedron. In the case of the 2-d {\it
kagom\'e} lattice, where the Ising axes now point towards the
centre of the triangular motif, the ice rules are that either 2
spins enter and one leaves each triangle, or the converse.
Inspection quickly reveals that there are 6 possible ground
state configurations associated
with each triangular motif. These are demonstrated in Fig.~\ref
{lattices_figures}a). Extension of the motifs in 2 dimensions shows that {\it
kagom\'e} spin ice possesses an extensive degeneracy, the residual entropy
of which is calculated in Section \ref{Residual entropy}.

\section{Residual entropy \`a la Pauling}
\label{Residual entropy}

The residual entropy of cubic ice, and thus so
of pyrochlore spin ice, has been evaluated by
Pauling\cite{Pauling} as $S_{p}= \frac{Nk}{2}ln3/2$. While this
value is very close to the exact one evaluated
numerically,\cite{Nagle} we note that in fact Pauling's
estimation is that of a Bethe lattice of tetrahedra: starting from
an arbitrarily chosen `central' tetrahedron, 6 of the $2^4=16$
possible spin states correspond to spin ice configurations. As the
4 neighboring tetrahedra each have 1 spin fixed due to sharing
with the central tetrahedron, they each have only 3 possible ice
states, from a complete manifold of $2^3=8$ states. The same
argument can then be used for all tetrahedra as one moves out
through the Bethe lattice. Thus, among the 2$ ^{N}\ $possible
states (N being the total number of sites), only a fraction $
\left( \frac{3}{8}\right) ^{N_{T}}$\ (N$_{T}$ being the number of
tetrahedra, i.e. N$_{T}$ = N/2) satisfy the ice rule on all
tetrahedra, which leads to an entropy:  $S_{p}=kln(2^{N}\times
\left( \frac{3}{8}\right) ^{N_{T}})=\frac{Nk}{2}ln3/2$. In the pyrochlore 
lattice, closed loops have to be taken into account: for one tetrahedron 
with 2 spins fixed by the neighboring tetrahedra, the number of acceptable 
ice states is either 1 or 2 (depending on the configuration of the 2 fixed spins), 
among 4 possible configurations for the 2 remaining spins; this leads again 
to an average reduction factor of $\frac{3}{8}$ , while if 3 spins are fixed 
this factor is larger: $\frac{1}{2}$. Thus for a lattice of tetrahedra, the 
number of ice states should be slightly larger than the Pauling value $S_{p}$, 
as has been found numerically $S_n=1.01S_p$.\cite{Nagle}. As previously 
demonstrated for more general ice models \cite{LiebWu}, Pauling's 
estimate is a lower bound on the exact value of the entropy.

The estimation of the residual entropy for a Bethe lattice of
triangles (Fig.~\ref{lattices_figures}b) can be made following the
same principles: for each triangle with 1 fixed spin, the number
of ice states is equal to 3, while the number of possible states
is $2^{2}=4$.\ Thus, if N$_{T}$ is now the number of triangles,
the total number of ice states is $2^{N}\times \left(
\frac{3}{4}\right) ^{N_{T}}$, (with $N_{T}=\frac{2N}{3}$), leading
to an entropy $S_{b}=\frac{Nk}{3}ln9/2$. If we again consider the
effect of the closed loops present in the {\it kagom\'{e}} lattice 
(Fig.~\ref{lattices_figures}c), we find that if 2 spins are fixed for 
a given triangle, the number of acceptable states for the third spin 
is either 1 or 2, among 2 possible states. Assuming that the states 
with 2 spins fixed are equiprobable, an average factor $\frac{3}{4}$ 
is again obtained. Although this procedure ignores any correlations beyond the 
first neighbors, this simplified argument suggests that 
closed loops should not be expected to change that much
the entropy in the {\it kagom\'{e}} lattice and the exact value
should be very close to $S_{b}$. Our numerical estimations show
clearly that this is the case. In fact this spin ice 
{\it kagom\'{e}}  system can be mapped exactly on the Ising 
{\it kagom\'{e}} antiferromagnet, in a similar way as the pyrochlore
system.\cite{Liebmann1,Harris97} Consequently, the value that we
found for the entropy is the same that was found for disordered ground state of the
AF Ising {\it kagom\'e} system.\cite{Liebmann2,Huse}


\section{Monte Carlo calculations}
\label{Monte Carlo calculations}

Our classical\cite{Quantum_fluctuations}  simulations were made on
a single {\it kagom\'{e}} plane of $50\times 50$ unit cells, each
containing 3 spins of modulus $S=1$, i.e. 7 500 spins in total.
The nearest- and second neighbor exchange integrals, $J_1$ and $J_2$,
were used to couple the spins according to the pathways shown in
Fig.~\ref{lattices_figures}d. A Kirkpatrik\cite{Kirkpatrick_MC} cooling scheme was used
with each step in temperature involving a reduction from $T_{1}$
to $T_{2}$ according to the equation $ T_{2}=0.9T_{1}$. At each
temperature 5 000 cycles site$^{-1}$ were used for equilibration
and 5 000 cycles site$^{-1}$ for the calculation of the
thermodynamic properties. The starting temperature for all the
simulations was $T=10|J_{1}|S^2$. The spin maps presented are
representative of the average over the Monte Carlo cycles
performed at each temperature that are within the specified range.

The Hamiltonian can be defined as:

\begin{equation}
H = - \sum\limits_{i,j < i} J_{ij} \vec {S}_i \cdot \vec {S}_j + H_A
- \left( g\mu _B \right)\vec {H}\cdot \sum\limits_i \vec {S}_i
\label{equation_1}
\end{equation}

\noindent where the anisotropy term $H_A$ reads $\sum\limits_{i}
\kappa_{i} \left( \vec{S}_i \cdot \hat{x}_i \right)^2$.
$\kappa_{i} < 0$ is the single-ion anisotropy constant that forces
the local spin $\vec{S}_i$ to align along the local axis
$\hat{x}_i$ and $\vec{H}$ is the applied field. In order to create Ising spins with the different
local orientations shown in Fig.~\ref{lattices_figures}, 
the transverse spin fluctuations were completely frozen out, 
which corresponds to setting $\kappa_{i}=-\infty$.

\begin{figure}[tbp]
\centerline{\hbox{\epsfig{figure=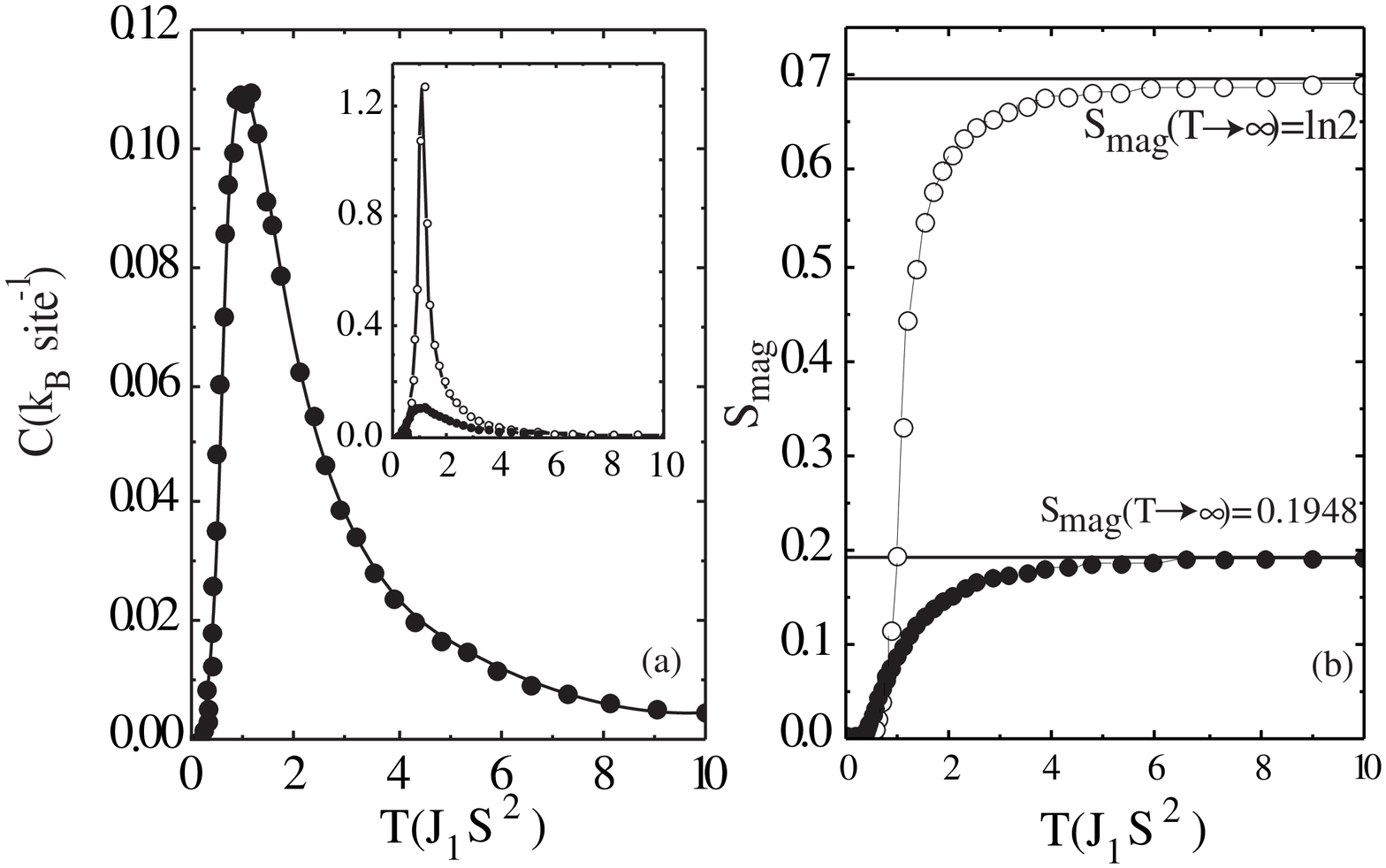,width=8cm}}}
\caption{(a) The specific heat and the (b) magnetic entropy  of
{\it kagom\'e} spin ice ($\bullet $) and the related
antiferromagnetic system ($\circ $) as a function of temperature.
Due to their differing scales, the comparison between the specific
heat of the two systems is shown as an insert. The limiting values
of their magnetic entropies, $S_{mag}$, as $T\rightarrow\infty$
are shown.} \label{entropy_figure}
\end{figure}

\section{Thermodynamic properties and spin correlations of kagom\'e spin ice}
\label{Thermodynamic properties of kagome spin ice}

The specific heat of {\it kagom\'{e}} spin ice shows a 
broad maximum at $T/J_{1}S^2=1$. Above
this temperature, in the range $1<T/J_{1}S^2<2$, the slow approach of
the magnetic entropy to its saturation level reveals the presence
of significant short-ranged order correlations. The residual
entropy of the spin ice state is clearly discernible in the
magnetic entropy (Fig.~\ref{entropy_figure}) as the shortfall in
the high temperature limit with respect to the $ln2$ expected for
Ising spins as $T\rightarrow\infty $. If the limiting value of the
integrated entropy is taken as $0.1948~(35)$, we find that the residual
entropy of the spin ice phase is $S_{mag}(T=0)=S_{mag}(T=\infty
)-\int_{0}^{\infty }\frac{C}{T} dT=ln2-0.1948=0.4982~(35)$, a value
that is in good agreement with the residual entropy of
$S_{b}=\frac{1}{3}ln9/2=0.5014$ calculated for the Bethe lattice.

\begin{figure}[tbp]
\centerline{\hbox{\epsfig{figure=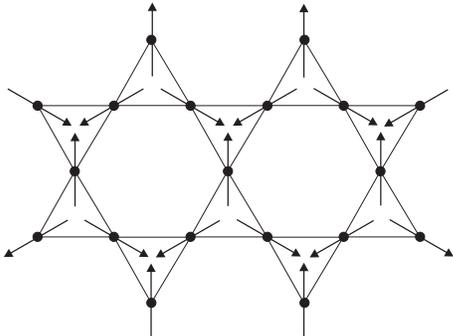,width=6cm}}}
\caption{The `q=0' ground state of the {\it kagom\'e} antiferromagnet.} 
\label{q=0_figure}
\end{figure}

As a contrast, the entropy of the antiferromagnetic system
($J_1<0$) with the same Ising anisotropies as the spin ice
phase is also shown in Fig.~\ref {entropy_figure}. While a maximum
in the specific heat is also seen at $T_C= |J_1S^2|$ (Fig.~\ref
{entropy_figure}a), the magnetic entropy differs sharply from the
{\it kagom\'e} spin ice in that it involves a recovery of the
complete $S=ln2$ value expected for the formation of long-range
order. The long-range order that results is the so-called {\it
q=0} spin structure familiar in {\it kagom\'e}
antiferromagnets\cite{Harris92} and is drawn in Fig.~\ref{q=0_figure}.

Equal time correlations $<\vec{S}_i \cdot \vec{S}_j>$ between distant 
spins $\vec {S}_i$ and $\vec{S}_j$, averaged over the Monte Carlo cycles 
at equilibrium, were computed at different temperatures. 
Fig.~\ref{spin spin correlations figure} show the thermal variations 
obtained for first and second spin neighbors. An increase 
in absolute values is observed as the temperature is decreased, 
down to $T\sim 0.5|J_{1}|S^2$ at which the correlations saturate.

\begin{figure}[tbp]
\centerline{\hbox{\epsfig{figure=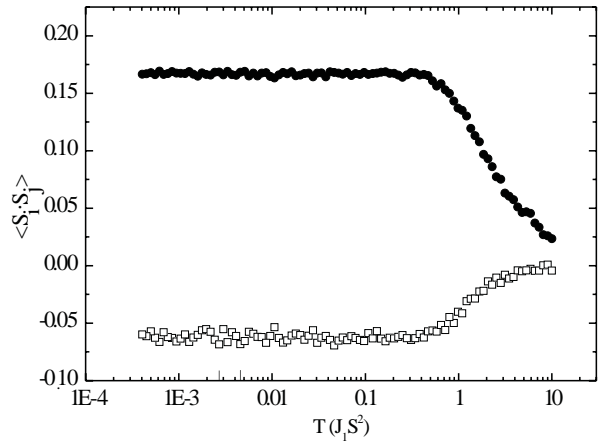,width=8cm}}} \caption{A
semi-log plot of the first ($\bullet $) and second neighbor
({\tiny$\square$}) spin-spin correlations in {\it kagom\'e}
spin ice as a function of temperature.} \label{spin spin
correlations figure}
\end{figure}

\begin{figure}[tbp]
\centerline{\hbox{\epsfig{figure=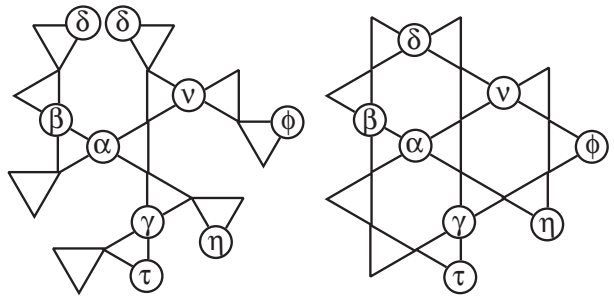,width=8cm}}}
\caption{Neighborhood of a given spin $\vec{S_{\alpha}}$ in the Bethe
and the {\it kagom\'e} lattice of triangles.}
\label{neighbor_bethe_kagom}
\end{figure}

The correlations at zero temperature in the spin ice phase are evaluated analytically for the Bethe lattice of triangles for all
neighbors as $<\vec{S}_i \cdot \vec{S}_j>_{b}=
\left( \frac{-1}{3}\right) ^{m} \cos \theta_{ij}$, where $m$ is 
the number of bonds that form the path joining the spins 
$\vec {S}_i$ and $\vec{S}_j$, and $\theta_{ij}$ is the angle 
difference $\theta_{j} - \theta_{i}$ between the orientations 
of the local anisotropy axes $\hat{x}_i$ and $\hat{x}_j$ for 
these spins. Numerical values for different neighbors 
$\vec{S_{\beta}}$, $\vec{S_{\gamma}}$, \ldots~of a given spin 
$\vec{S_{\alpha}}$ as specified in Fig.~\ref{neighbor_bethe_kagom} are 
reported in Table~\ref{table1}, and compared to the correlations for 
corresponding neighbors in the {\it kagom\'{e}} lattice, that were computed at the
lowest temperature ($T\sim 0.0004~J_{1}S^2$) of the Monte Carlo 
simulations. Agreement is remarkable, except for the 
$\vec{S_{\alpha}} - \vec{S_{\delta}}$ spin pairs. This is to be expected as 
in the {\it kagom\'e} lattice these neighbors belong to the same hexagon 
and are thus connected by two paths with exactly the same number
of bonds, which leads to correlations that are larger by a factor of 
$\sim 2$. In a similar way the other distant spins in the 
{\it kagom\'{e}} lattice, such as the second neighbors, are also not 
connected by only one path, however as these other paths have a larger 
number of bonds and the concomitant corrections decay as a power law 
in the number of bonds, they give rise to only minor adjustments.

\begin{table}
\begin{tabular}{cccccc}
$i~j$ & $m$ & $\theta_{ij}$ & $<\vec{S}_i \cdot \vec{S}_j>_{b}$ & 
$n^{th}$ & $<\vec{S}_i \cdot \vec{S}_j>$ \\
\hline
$\alpha~\beta$&$1$&$\pm2\pi/3$&$1/6=0.16667$&$1^{st}$&$0.166$ \\
$\alpha~\gamma$&$2$&$\pm2\pi/3$&$-1/18=-0.05556$&$2^{nd}$&$-0.059$\\
$\alpha~\nu$&$2$&$0$&$1/9=0.11111$&$3^{rd}$&$0.101$\\
$\alpha~\delta$&$3$&$0$&$-1/27=-0.03704$&$3^{rd}$&$-0.072$\\
$\alpha~\tau$&$3$&$\pm2\pi/3$&$1/54=0.01754$&$4^{th}$&$0.012$\\
$\alpha~\eta$&$3$&$\pm2\pi/3$&$1/54=0.01754$&$5^{th}$&$0.017$\\
$\alpha~\phi$&$4$&$0$&$1/81$=0.01234&$6^{th}$&$0.007$\\
\end{tabular}
\caption{Spin-spin correlations in the Bethe and {\it kagom\'e}
lattice of triangles between a given spin $\vec{S_{\alpha}}$ and its
neighbors as specified in Fig.~\ref{neighbor_bethe_kagom}
($n^{th}$: order of neighborhood in the {\it kagom\'e} lattice).}
\label{table1}
\end{table}

\section{Antiferromagnetic second neighbor interactions}
\label{Antiferromagnetic second neighbor interactions}

\begin{figure}[tbp]
\centerline{\hbox{\epsfig{figure=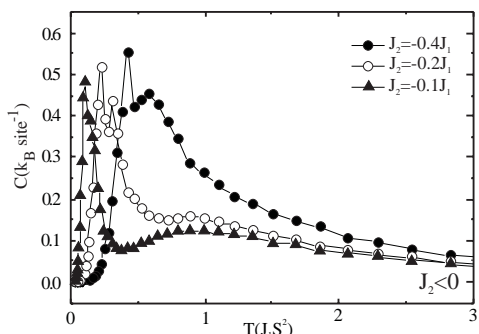,width=8cm}}}
\caption{Specific heat for the Ising {\it kagom\'{e}} ferromagnet
with second neighbor interactions: $J_{2}<0$.}
\label{specific_heat_AF}
\end{figure}

\begin{figure}[tbp]
\centerline{\hbox{\epsfig{figure=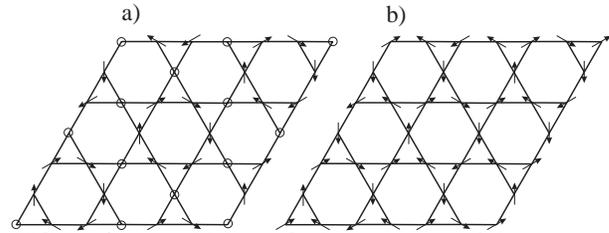,width=8cm}}}
\caption{Averaged Monte Carlo spin maps of the structures for
$J_{2}<0$ when a) $|J_{2}|S^2<T<1.5|J_{2}|S^2$  b)
$T<<|J_{2}|S^2$. Open circles represent disordered
spins.}\label{af_spin_configs}
\end{figure}

These were examined with the values of $J_{2}/J_{1}$ = -0.4, -0.2
and -0.1. In all cases 3 features are seen in the specific heat
(Fig.~\ref{specific_heat_AF}). The first at $T\sim J_{1}S^2$
involves the formation of short-ranged spin ice-type correlations.
Below this temperature, at $T\sim 1.5|J_{2}|S^2$, the additional
antiferromagnetic exchange stabilizes a partially ordered
structure that is made up of a triangular motif of clockwise and counter-clockwise
hexagons of spins (Fig.~\ref{af_spin_configs}a). The mean
magnitude of the moments on the sublattice that separate these
hexagons is zero. The reason for some of the sites failing to
order is made clear from examination of the lowest maximum at $
T\sim |J_{2}|S^2$: this involves the formation of an ordered
magnetic structure which has a unit cell that is $3\times $ larger
along each of the axes than the paramagnetic cell
(Fig.~\ref{af_spin_configs}b). The complexity of this structure is
evidenced from the 2 types of hexagon: the clockwise hexagons (and
the time-reversed counter-clockwise) are separated by those that have 3
pairs of parallel spins. As the latter involve adjacent spins
with antiferromagnetic 120$^{\circ }$ orientations this ground
state configuration is frustrated. Re-examination of the
intermediate phase ($|J_{2}|S^2<T<1.5|J_{2}|S^2$) shows that the
fluctuations have selectively annulled the mean moment on sites
that would otherwise frustrate the lattice.

As $J_2/J_1$ is increased, the intermediate phase becomes stable
over a wider temperature range; when $J_2/J_1=-0.4$ this phase
occurs at temperatures from 0.4$J_1 S^2$ to 0.6$J_1 S^2$.

\section{Ferromagnetic second neighbor interactions}
\label{Ferromagnetic second neighbor interactions}

\begin{figure}[tbp]
\centerline{\hbox{\epsfig{figure=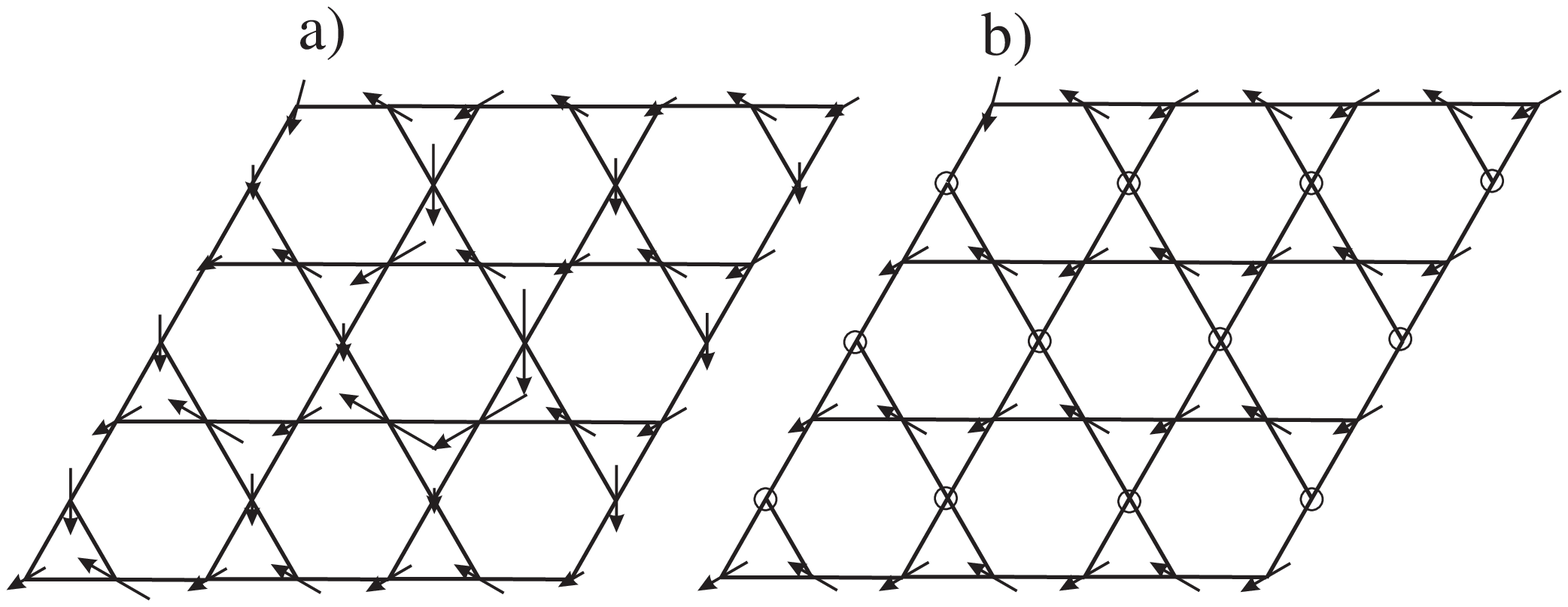,width=8cm}}}
\caption{While no long-range order occurs, tendencies towards
order with particular structures are seen. These spin maps are
presented for $J_2\gtrsim 0$ when a) $T\lesssim J_{2} S^2$ b)
$T<<J_2S^2$. Open circles represent disordered spins.}
\label{f_spin_configs}
\end{figure}

\begin{figure}[tbp]
\centerline{\hbox{\epsfig{figure=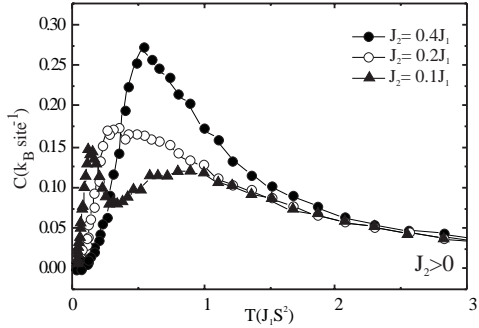,width=8cm}}}
\caption{Specific heat for the Ising {\it kagom\'{e}} ferromagnet
with second neighbor interactions: $J_{2}>0$.}
\label{specific_heat_F}
\end{figure}

As the lattice of second neighbors is also a  {\it kagom\'{e}}
lattice, ferromagnetic $J_{2}$ still result in a highly frustrated
spin system and if the moments related by $J_{1}$ and $J_{2}$ are
of uniform magnitude 1/3 of the moments lie on sites of zero
mean-field (Fig.~\ref{f_spin_configs}). It is not surprising,
therefore, that extensive fluctuations are still present at the
lowest temperatures for all the values of $J_{2}>0$ studied
($J_{2}/J_1=$ 0.4, 0.2, and 0.1), in sharp contrast with the ground
state when $J_{2}<0$. If these second neighbor interactions present a
minor contribution to the Hamiltonian, e.g. $J_{2}/J_1=0.1$,
short-ranged ice-like correlations still form and these give rise
to a broad feature in the specific heat at $T\sim J_{1}S^2$
(Fig.~\ref {specific_heat_F}). However, in contrast with the
$J_{2}<0$ phases, only one maximum is seen at lower temperature,
$T\sim J_{2}S^2$. Remarkably, this is not associated with a 
phase transition and no long-range order is seen down to the lowest
temperature: only a tendency is observed that involves the condensation
of a partially ordered phase with the propagation vector {\bf
k}=(0 0) and a non-zero ferromagnetic component
(Fig.~\ref{f_spin_configs}). These correlations are strongest when $J_2$ is
weakest. As $J_2$ increases, this tendency dwindles out and the two
maxima in the specific heat merge. What remains are strongly
fluctuating phases with correlations that are less and less
reminiscent of the partially ordered phase.\cite{Ballou}

\section{Effects of magnetic field}
\label{Effects of field}

\begin{figure}[tbp]
\centerline{\hbox{\epsfig{figure=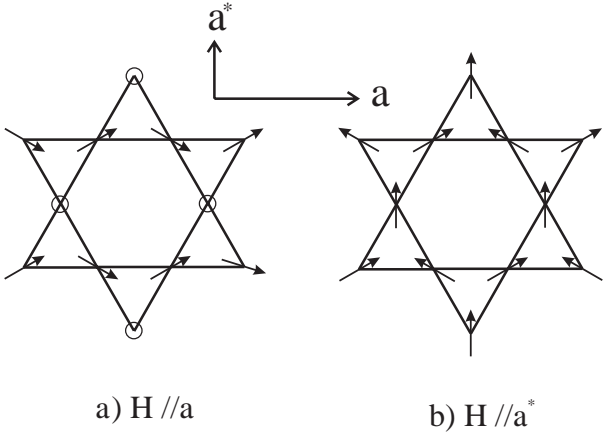,width=8cm}}}
\caption{The high field spin configurations when the applied field
is along the ${\bf a}$ and ${\bf a^*}$ directions. Open circles
represent disordered spins, i.e. those that have equal probability
of being in either direction.} \label{field_sweep_structures}
\end{figure}

The effects of an externally applied magnetic field on the spin ice
phase were studied with fields of up to 0.269~$J_1S$ applied along
the characteristic directions of the {\it kagom\'e} lattice, {\it
i.e.} ${\bf a}$ and ${\bf a^*}$
(Fig.~\ref{field_sweep_structures}). These calculations were made at
a temperature of 0.025~$J_1S^2$ after the system had beencooled in
zero applied field.

\begin{figure}[tbp]
\centerline{\hbox{\epsfig{figure=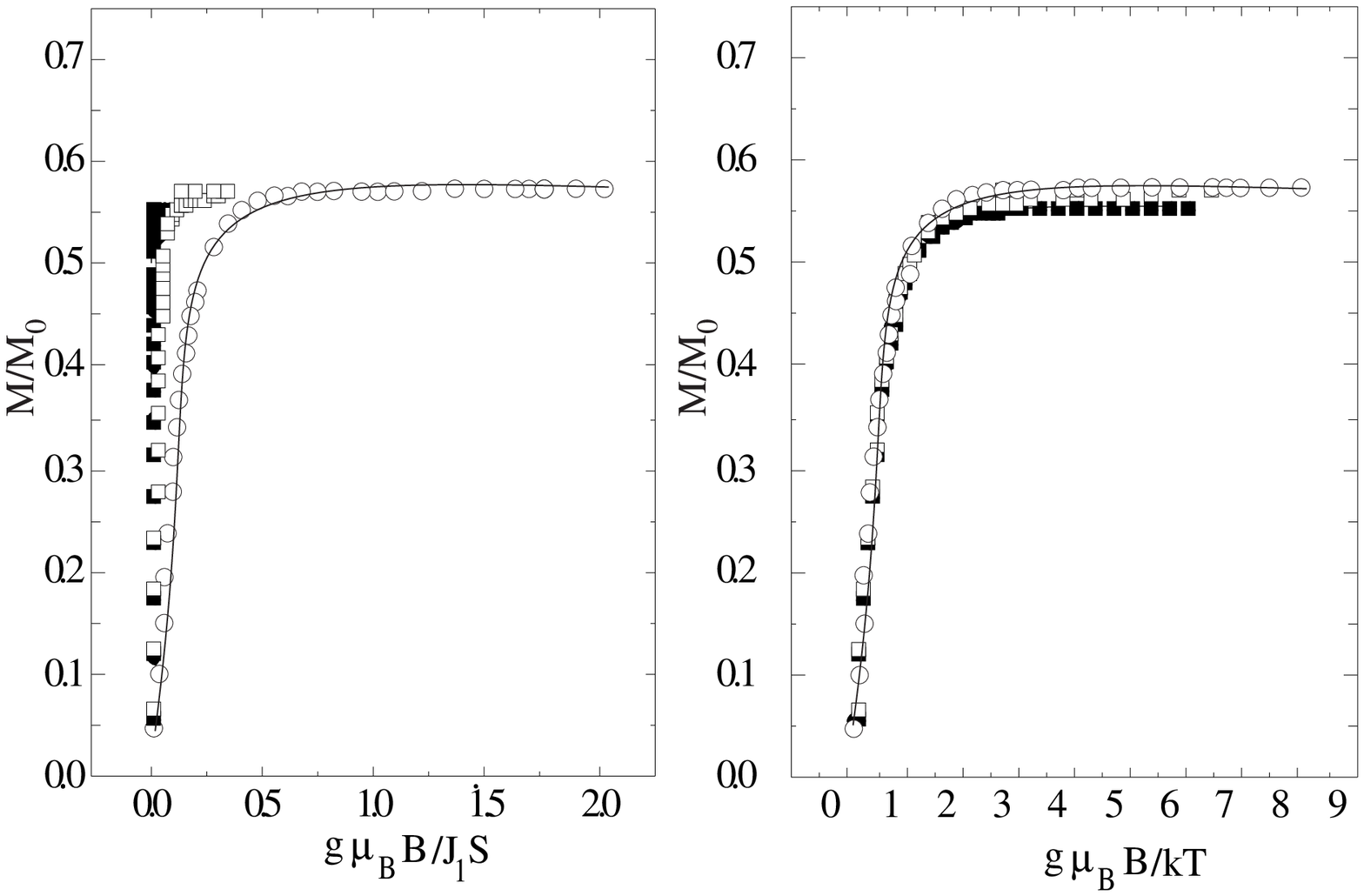,width=8cm}}}
\caption{The reduced magnetization per site as a function of applied
field along the ${\bf a}$ direction (a) in units of the molecular
field $J_1S$ and (b) after rescaling the applied field with
temperature, i.e. plotting $M/M_0$ {\it vs.} $\mu_B B g/kT$, a
universal curve is obtained. The temperatures of the field
sweeps were ($\circ$) 0.254, ({\tiny$\square$}) 0.051, ($\bullet$)
0.010, and ({\tiny$\blacksquare$}) 0.002~$J_1S^2$. The lines provides
guides to the eye.} \label{field_sweep_along_a}
\end{figure}

With the field parallel to the ${\bf a}$ direction a continuous
increase in an ordered component is observed. Notably,
it involves only 2/3 of the sites (Fig.~\ref{field_sweep_structures}a). 
This demonstrates another point
of similarity between the 2-d {\it kagom\'e} spin ice system and
that of the closely related 3-d pyrochlore lattice: the spin
components induced by a field along some of the crystallographic
directions are incapable of choosing a unique ground state and
significant degeneracy is retained. A common result is the formation of  
only partially-ordered spin configurations.

With the field parallel to the ${\bf a^*}$ direction a continuous
increase of order with the {\bf k}=(0 0) structure
(Fig.~\ref{field_sweep_structures}b) is again found. Trivially we find
that the ordering is preferential for the sites with their local
anisotropy parallel to the field direction. The ordered component
of all the sites saturates in a field of 0.054~$J_1S$.

\begin{figure}[tbp]
\centerline{\hbox{\epsfig{figure=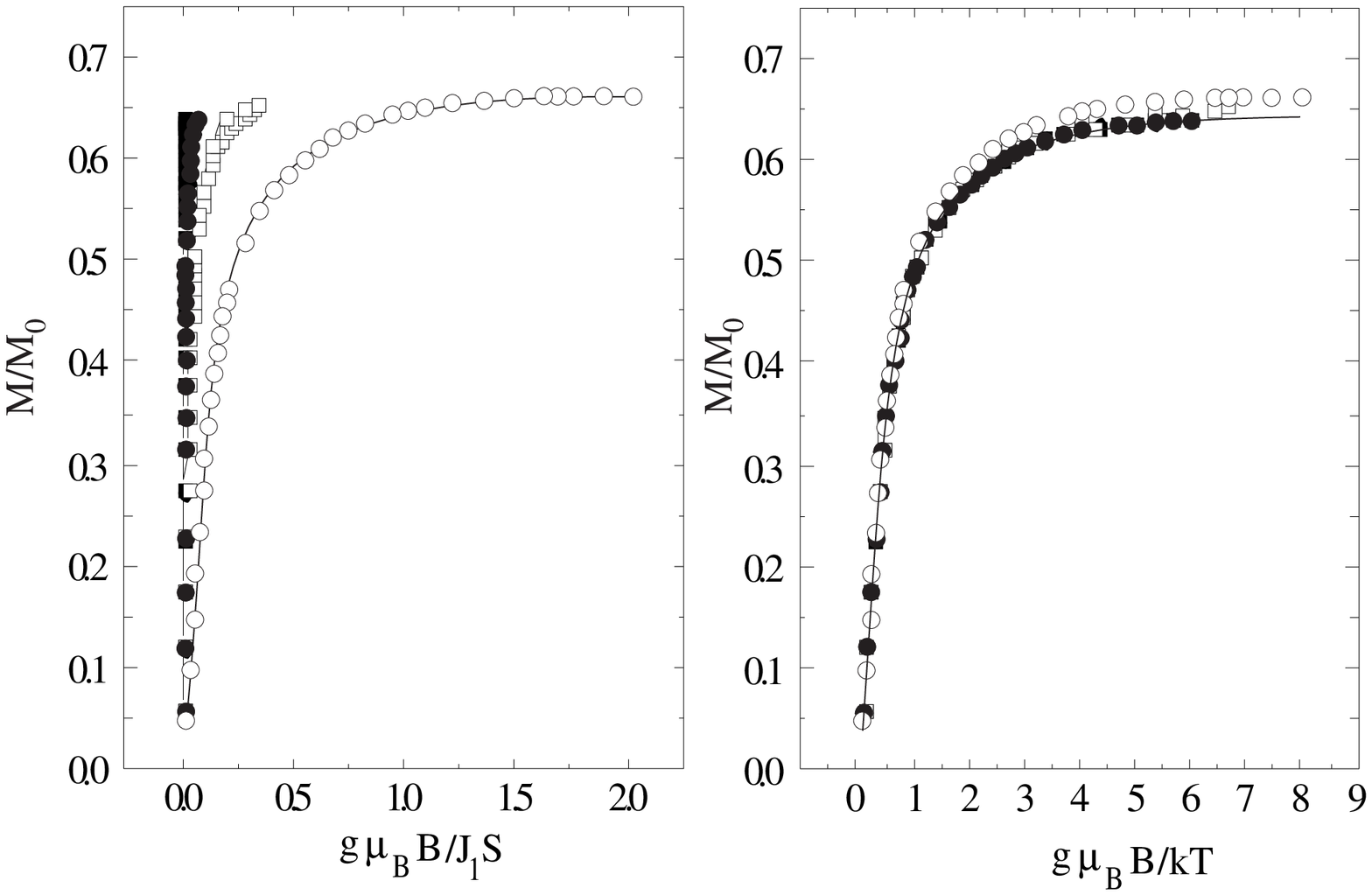,width=8cm}}}
\caption{The reduced magnetization per site as a function of
applied field along the ${\bf a^*}$ direction (a) in units of the
molecular field $J_1S$ and (b) after rescaling the applied field
with temperature, i.e. plotting $M/M_0$ {\it vs.} $\mu_B B g/kT$,
a universal curve is obtained. The temperatures of the field
sweeps were ($\circ$) 0.254, ({\tiny$\square$}) 0.051, ($\bullet$)
0.010, and ({\tiny$\blacksquare$}) 0.002~$J_1S^2$. The lines
provides guides to the eye.} \label{field_sweep_along_a_star}
\end{figure}

Inspection of the high-field ordered structures suggests that even an
infinitesimal field applied along either of the characteristic directions would induce an ordered component. It is therefore
a surprise that the initial dependence of the reduced magnetization on
the field is not sharper. Our simulations show that this weaker dependence
is a result of thermal fluctuations, rather than being a consequence of
the finite lattice size. This is demonstrated by the collapse of the
reduced magnetization as a function of the Zeeman energy divided by the
thermal energy, onto the same curve at all temperatures studied 
(see Fig.~\ref{field_sweep_along_a} and 
Fig.~\ref{field_sweep_along_a_star}).

\section{Discussion}
\label{Discussion}

The thermodynamic properties of {\it kagom\'e}
spin ice, and those of phases in the presence of second neighbor
interactions and applied fields, manifest clearly the richness of
frustrated physics once thought to be the domain only of
antiferromagnets. In particular, the importance of spin
fluctuations is central and their presence leads to the
stabilization of a variety of magnetic structures with both
disordered and ordered sublattices. While such partially ordered
phases have already found to be stable in frustrated lattices when
the local moments are close to an
instability\cite{partial_ordering} or in systems with contrived
further-neighbor exchange,\cite{partial_ordering_2} here they are
stable with well defined magnetic moments and are a simple
consequence of the strong frustration associated with uniaxial
anisotropy. In real systems\cite{Note added to proof} additional terms such as quantum fluctuations and dipolar interactions will be present in the Hamiltonian. While further work is required to determine how these will alter
the spin ice phase and the partially ordered phases presented here,
these effects have already been shown to be strongly influential on
their low-temperature physics. A review of how Ising models are influenced by the introduction of quantum dynamics that arise from the inclusion of an $XY$ exchange component is presented in Ref.~\onlinecite{Moessner_Ising_quantum_fluctuation}. An interesting limit for their inclusion is the $XY$ antiferromagnetic on a pyrochlore lattice model presented by Champion et al.\cite{Champion_XY}, in which spin-wave analysis indicated that zero-point quantum fluctuations stabilize the formation of an ordered moment from the a highly degenerate continuous manifold.

The observation that the ordered N\'eel state proposed\cite{Wills-PRB-Jarosites} for the vanadium jarosite ${\rm {(H_3O)V_3(SO_4)_2(OH)_6}}$ is a highly symmetric member of the spin-ice manifold suggests that spin ice-like correlations will be present above, but close to, its ordering temperature. The system is clearly not a true realization of {\it kagom\'e} spin ice as the presence of a transition to N\'eel order requires that either additional terms in the exchange Hamiltonian that we have not considered here have raised the degeneracy of the ice-like manifold, or that the single-ion anisotropies it possesses are of a magnitude that quantum fluctuations are important. Unfortunately, the single-ion terms in the jarosite are  difficult to estimate because the relative contributions of axial
$(DS^2_z)$ and rhombic $(E(S^2_x-S^2_y))$ terms determined from
magnetic susceptibility
data\cite{Papoutsakis} taken from  ${\rm {KGa_{2.96}V_{0.04}(SO_4)_2(OH)_6}}$ are ambiguous, as other pairs $(D,E)$ parameters may fit them equally well. It is still, however, reasonable that the short-range correlations present above the ordering temperature will be characteristic of the most important terms in the Hamiltonian, and thus of the spin ice phase.

In conclusion, we have presented a new model of localized highly
frustrated ferromagnetism: {\it kagom\'e} spin ice. If residual
entropy is used as a measure of frustration, our analytical and
numerical calculations demonstrate that {\it kagom\'e} spin ice is
more highly frustrated than the 3-d counterpart, pyrochlore spin
ice. It is therefore the most highly geometrically frustrated
``ferromagnetic'' ground state yet studied. Furthermore,
second neighbor interactions within the {\it kagom\'e}
plane are shown to have the salient effect of stabilizing
structures that feature both ordered and disordered sublattices.


\begin{references}
\bibitem{Quantum_fluctuations}
We note that quantum fluctuations cannot play a r\^ole in the
physics of spin ice systems at low temperature since these states
require large Ising anisotropies that will freeze them out.
\bibitem{Note added to proof}
Since the submission of this manuscript a kagom\'e spin ice phase has been stabilized in $\rm{Dy_2Ti_2O_7}$ by the application of a magnetic field along the $<111>$ direction.\cite{Matsuhira}
\end{references}
\end{document}